\documentclass[12pt,oneside,english]{amsart}
\usepackage{lmodern}

\usepackage{geometry}
\usepackage{color}
\usepackage{babel}
\usepackage{rotfloat}
\usepackage{booktabs}
\usepackage{amstext}
\usepackage{amsthm}
\usepackage{amssymb}
\usepackage{graphicx}
\usepackage{setspace}
\usepackage{bbold}

\usepackage[authoryear]{natbib}
\usepackage[unicode=true,pdfusetitle,
 bookmarks=true,bookmarksnumbered=false,bookmarksopen=false,
 breaklinks=false,pdfborder={0 0 0},pdfborderstyle={},backref=false,colorlinks=true]
 {hyperref}
\hypersetup{
 citecolor=blue, linkcolor=blue}
\usepackage{breakurl}

\makeatletter

\theoremstyle{plain}
\newtheorem{thm}{\protect\theoremname}
  \theoremstyle{remark}
  
  \theoremstyle{plain}
  
  \theoremstyle{plain}
  
  \theoremstyle{plain}
  \newtheorem{lem}{\protect\lemmaname}

\usepackage{calc}\usepackage{fixltx2e}\usepackage{multicol}\usepackage[normalem]{ulem}
\usepackage{geometry}
\usepackage{array}
\usepackage{float}
\usepackage{multirow}
\usepackage{amstext}
\usepackage{amsthm}
\usepackage{amssymb}
\usepackage{graphicx}
\usepackage{setspace}
\usepackage{esint}
\usepackage{graphicx,amsmath,amssymb,epsfig}
\usepackage{amsfonts}
\usepackage{geometry}
\usepackage{setspace}
\usepackage{lscape}
\usepackage{caption}
\usepackage{rotating}
\usepackage{xcolor}
\usepackage{amsfonts}
\usepackage{subfig}
\usepackage{babel}

\theoremstyle{definition}\newtheorem{assumption}{Assumption}

\makeatletter
\floatstyle{ruled}
\theoremstyle{remark}
  
\thanks{\noindent $^\dag$ Department of Economics, MIT, 
wnewey@mit.edu.}
\thanks{ $\S$ Department of Economics, University of Bristol, 
s.stouli@bristol.ac.uk.}
 
\@ifundefined{showcaptionsetup}{}{%
 \PassOptionsToPackage{caption=false}{subfig}}
\usepackage{subfig}
\makeatother

\providecommand{\lemmaname}{Lemma}
\providecommand{\propositionname}{Proposition}
\providecommand{\remarkname}{Remark}
\providecommand{\corollaryname}{Corollary}
\providecommand{\theoremname}{Theorem}

\usepackage{tikz}
\usetikzlibrary{arrows,shapes.arrows,shapes.geometric,
	shapes.multipart,backgrounds,decorations.pathmorphing,positioning,fit,automata}
\tikzset{
	>=stealth',
	sq/.style={
		rectangle,
		draw=black, very thick,
		minimum height=2.8em,
		minimum width=2.8em,
		text centered},
	sqf/.style={
		rectangle,
		draw=black, very thick,text=black,
		minimum height=2.8em,
		minimum width=2.8em,
		text centered,fill=gray!80, opacity = 0.8},
	est/.style={
		circle,
		draw=black, very thick,
		minimum height=3em,
		minimum width=3em,
		text centered},
	myempty/.style={
	circle,
	draw=white, very thick,
	text centered},
}

\begin{document}

\title{Heterogeneous Coefficients, Control Variables, and Identification of Multiple Treatment  
Effects}

\author{Whitney K. Newey$^\dag$ and Sami Stouli$^\S$}
\begin{abstract}
Multidimensional heterogeneity and endogeneity are important features
of models with multiple treatments. We consider a heterogeneous coefficients
model where the outcome is a linear combination of dummy treatment
variables, with each variable representing a different kind of treatment.
We use control variables to give necessary and sufficient conditions
for identification of average treatment effects. With mutually exclusive
treatments we find that, 
provided the heterogeneous coefficients
are mean independent from treatments given the controls, a simple
identification condition is that the generalized propensity scores
(\citealp{Imbens:2000}) be bounded away from zero and that their
sum be bounded away from one, with probability one. 
Our analysis extends to distributional and quantile treatment effects, as well as corresponding treatment effects on the treated. These
results generalize the classical identification result of \citet{RR:1983}
for binary treatments. 
\end{abstract}

\maketitle
\textsc{Keywords}: Treatment effect; Multiple treatments; Heterogeneous coefficients; Control variable; Identification; Conditional nonsingularity; Propensity score.\

\section{Introduction}

Models that allow for multiple treatments are important for program
evaluation and the estimation of treatment effects (\citealp{Catt:2010,Imai vD:2004,Imbens:2000,Graham Pinto 2018,Lechner:2001}). 
A general class is heterogeneous coefficients models where the outcome
is a linear combination of dummy treatment variables and unobserved
heterogeneity. These models allow for multiple treatment regimes, with
each dummy variable representing a different kind of treatment. These models
also feature multidimensional heterogeneity, with the dimension of
unobserved heterogeneity being determined by the number of treatment
regimes.

Endogeneity is often a problem in these models because we are interested
in the effect of treatment variables on an outcome, and the treatment
variables are correlated with heterogeneity. Control variables provide
an important means of controlling for endogeneity with multidimensional
heterogeneity. For treatment effects, a control variable is an observed
variable that makes heterogeneity and treatment variables independent when it
is conditioned on (\citealp{RR:1983}).

We use control variables to give necessary and sufficient
conditions for identification of average treatment effects based on
conditional nonsingularity of the second moment matrix of the vector
of dummy treatment variables given the controls. This first main result is familiar
in the binary treatment case, but its generalization to multiple treatments 
appears to be new. With mutually exclusive treatments
we find that, 
provided the heterogeneous coefficients
are mean independent from treatments given the controls, a simple
identification condition is that the generalized propensity scores
(\citealp{Imbens:2000}) be bounded away from zero and that their
sum be bounded away from one, with probability one. This condition is the same as common support, 
that the support of treatment variables conditional on the controls is equal to the marginal support of the treatment variables. 
Thus our second main contribution is to show that, with mutually exclusive treatments, conditional mean independence and 
common support are jointly sufficient for identification, 
a substantial weakening of the standard assumption that conditional independence
hold jointly with common support (e.g., \citealp{Frolich:2004}). 
We also extend our analysis to distributional and quantile treatment effects, 
as well as corresponding treatment effects on the treated. 
These results provide an important generalization of \citet{RR:1983}'s classical identification
result for binary treatments.

\section{Modeling of Treatment Effects}

\subsection{Modeling framework}

Let $Y$ denote an outcome variable of interest, and $D$ a vector
of dummy variables $D(t)$, $t\in\mathcal{T}\equiv\{1,\dots,T\}$,
taking value one if treatment $t$ occurs and zero otherwise, and
$\beta$ a structural disturbance vector of finite dimension.
We consider a heterogeneous coefficients model of the form
\begin{equation}
Y=p(D)^{\textrm{T}}\beta,\quad\quad p(D)=\{1,D(1),\dots,D(T)\}^{\textrm{T}}.\label{eq:g(d,b)}
\end{equation}
This model is linear in the treatment dummy variables, 
with coefficients
$\beta$ that are random and need not be independent of $D.$ 
%
%

When the potential outcome framework of \citet{Rubin:1974} is extended to mutually
exclusive treatment regimes in the definition of $D$, linearity in model (\ref{eq:g(d,b)}) arises naturally.
Denote the 
vector of potential outcomes by $\{Y(1),\ldots,Y(T)\}^{\textrm{T}}$,
and the potential outcome in the absence of treatment by $Y(0)$.
The observed outcome $Y$ and the vector of potential outcomes are
related by
\[
Y=Y(0)+\sum_{t=1}^{\textrm{T}}D(t)\{Y(t)-Y(0)\},
\]
which is of the form (\ref{eq:g(d,b)}) upon setting $\beta=(\beta_{0},\beta_{1},\ldots,\beta_{T})^{\textrm{T}}$
with  $\beta_{0}\equiv Y(0)$ and $\beta_{t}\equiv Y(t)-Y(0)$,
$t\in\mathcal{T}$. 

Mutually exclusive treatment regimes are important in a wide
variety of nonexperimental settings, such as program or policy evaluation
with a multivalued treatment (e.g., \citealp{Ao et al:2021,Lechner:2002,Uysal:2015}).
Consider for example evaluation of active labor market programs 
with a 
treatment taking on $T\geq2$ values, according
to different types or levels of program participation. 
Central objects
of interest are average effects on earnings for each treatment value,
\[
E(\beta_{t})=E\{Y(t)-Y(0)\},\quad t\in\mathcal{T}.
\]
Allowing for  multivalued treatments thus permits to capture different
average effects across program types or levels, going beyond the sole
effect of program participation considered in binary treatment analysis.
Another important example is policy or medical treatment evaluation
with non-mutually exclusive policies or treatments, implemented
both separately and jointly. In that case, a distinct treatment dummy variable is assigned
to each policy or treatment and to each implemented policy mix (\citealp{Becker Egger:2013,Tortu et al:2020}) or combined therapy (\citealp{Feng et al:2012,Nian et al:2019}).
Resulting treatment regimes in the definition of $D$ are then
mutually exclusive, by construction.

A main motivation
for defining the components of $D$ according to mutually exclusive
treatment regimes is the general validity of model (\ref{eq:g(d,b)})
in that case. In contrast, when treatment regimes are non-mutually
exclusive, model (\ref{eq:g(d,b)}) restricts the average effect of
any combination $\mathcal{C}$ of $K\leq T$ treatments $t_{1},\ldots,t_{K}\in\mathcal{T}$ 
to be additive in the average effects of each component of $\mathcal{C}$,
i.e., the average treatment effect of $\mathcal{C}$ is
\begin{equation}
\sum_{t\in\mathcal{C}}E(\beta_{t}) = \sum_{t\in\mathcal{C}}E\{Y(t)-Y(0)\},\quad\mathcal{C}=\{t_{1},\ldots,t_{K}\}. \label{eq:additiveTE}
\end{equation}
In addition to average effects of each treatment, model (\ref{eq:g(d,b)})
is then able to capture average
effects of potentially complex interventions by restricting the form
the average effects of combined treatments can take. The additivity restriction
has been used in evaluation of randomized medical experiments implementing
combinations of a large number of 
treatments (
see \citealp{Petropoulou et al:2021}, for a literature review), 
but does not appear to 
be common in nonexperimental settings.

In general, heterogeneous coefficients $\beta$ need
not be independent of $D$ because of confounding factors denoted
$X$. Here we assume that these factors are observable and that there
is sufficient independent variation in $D$ from $\beta$ once
conditioning on $X$. 
In the empirical examples above, $X$ includes a variety of individual characteristics of program participants, such as age, gender, measures of cognitive and non-cognitive skills, as well as socio-economic characteristics. 
Formally, we 
assume that
the vector $\beta$ is mean independent of the endogenous treatments
$D$, conditional on an observable control variable $X$.

\begin{assumption}For the model in (\ref{eq:g(d,b)}), there exists
a control variable $X$ such that $E(\beta\mid D,X)=E(\beta\mid X)$.
\label{ass:Assumption1} \end{assumption}

The \citet{RR:1983} treatment effects model is included as a special
case where $D\in\{0,1\}$ is a treatment dummy variable that is equal
to one if treatment occurs and equals zero without treatment, and 
\[
p(D)=(1,D)^{\textrm{T}}.
\]
In this case $\beta=(\beta_{0},\beta_{1})^{\textrm{T}}$ is two
dimensional with $\beta_{0}$ giving the outcome without treatment, 
and $\beta_{1}$ being the treatment effect. Here the control
variables in $X$ would be observable variables such that Assumption \ref{ass:Assumption1}
holds, i.e., the coefficients $(\beta_{0},\beta_{1})$
are mean independent of treatment conditional on controls; this is
the unconfoundedness assumption of \citet{RR:1983}.

\subsection{The average structural function\label{subsec:ASF}}

A central object of interest in model (\ref{eq:g(d,b)}) is the average
structural function given by $\mu(D)\equiv p(D)^{\textrm{T}}E(\beta)$;
see \citet{Cham:1984}, \citet{Blundell Powell 2003} and \citet{Wool:2005}. 
This function
is also referred to as the dose-response function in the statistics
literature (e.g., \citealp{Imbens:2000}). When $D\in\{0,1\}$ is
a dummy variable for treatment, $\mu(0)$ gives the average outcome
if every unit remained untreated and $\mu(1)$ the average outcome
if every unit were treated, with $\mu(1)-\mu(0)$ being the average
treatment effect. In general, the average effect of some treatment
$t\in\mathcal{T}$ is
\[
\mu(e_{t})-\mu(0_{T}),
\]
with $e_{t}=(0,\ldots,0,1,0,\ldots,0)^{\textrm{T}}$ defined as a $T$-vector
with all components equal to zero, except the $t$th, which is one,
and $0_{T}$ a $T$-vector of zeros.
Pairwise average treatment
effect comparisons are formed as $\mu(e_{t})-\mu(e_{s})$, for any
$s,t\in\mathcal{T}$, $s\neq t$. For non-mutually exclusive treatment  regimes,
the average effect of some combination $\mathcal{C}$ of treatments
$t_{1},\ldots,t_{K}\in\mathcal{T}$, $K\leq T$, is formed as
$\sum_{s\in\mathcal{C}}\left\{ \mu(e_{s})-\mu(0_{T})\right\} $, 
$\mathcal{C}=\{t_{1},\ldots,t_{K}\}$, 
and the corresponding relative average effect with respect to some
treatment $t\in\mathcal{T}$ as $\sum_{s\in\mathcal{C}}\left\{ \mu(e_{s})-\mu(e_{t})\right\} $.

The conditional mean independence assumption and the form of the structural
function $p(D)^{\textrm{T}}\beta$ in (\ref{eq:g(d,b)}) together imply
that the control regression function of $Y$ given $(D,X)$, $E(Y\mid D,X)$,
is a linear combination of the treatment variables:
\begin{equation}
E\left(Y| D,X\right)=p(D)^{\textrm{T}}E\left(\beta| D,X\right)=p(D)^{\textrm{T}}E\left(\beta| X\right)=p(D)^{\textrm{T}}q_{0}(X),\; q_{0}(X)\equiv E\left(\beta| X\right).\label{eq:controlregression}
\end{equation}
The average structural function can thus be expressed as a known linear
combination of $E\{q_{0}(X)\}$ from equation (\ref{eq:controlregression}).
By iterated expectations, 
\begin{equation}
p(D)^{\textrm{T}}E\left\{q_{0}(X)\right\}=p(D)^{\textrm{T}}E\left\{E\left(\beta\mid X\right)\right\}=\mu(D).\label{eq:mu(D)}
\end{equation}
We use the varying coefficient structure of the control regression
function (\ref{eq:controlregression}) and the implied linear form
of $\mu(D)$ to give conditions that are necessary as well as sufficient
for identification. For non-mutually exclusive treatment regimes, 
Appendix \ref{sec:Model} gives an example of a model for which the implied average structural function is of the linear form (\ref{eq:mu(D)}) while the average effect of some combination of treatments takes the additive form (\ref{eq:additiveTE}).

\section{Identification Analysis}

\subsection{Main results}

Under the maintained Assumption \ref{ass:Assumption1}, a sufficient condition for identification of the average structural
function is nonsingularity of the second moment matrix of the treatment
dummies given the controls,
\[
E\left\{p(D)p(D)^{\textrm{T}}\mid X\right\},
\]
with probability one. Under the additional assumption that $E\{p(D)p(D)^{\textrm{T}}\}$
 is nonsingular, this condition is also necessary. 
 
Theorem \ref{thm:Theorem1} states our first main 
result. The proofs
of all formal results are given in Appendix \ref{sec:Proofs}.
\begin{thm}
Suppose that $E(\left\Vert \beta\right\Vert ^{2})<\infty$,
$E\{p(D)p(D)^{\textrm{T}}\}$ is nonsingular, and Assumption \ref{ass:Assumption1}
holds. Then: $E\{p(D)p(D)^{\textrm{T}}\mid X\}$ is nonsingular with probability one
if, and only if, $\mu(D)$ is identified. \label{thm:Theorem1} 
\end{thm}

When $D\in\{0,1\}$ and $p(D)=(1,D)^{\textrm{T}}$, the identification
condition becomes the standard condition for the treatment effect
model 
\[
Y=\beta_{0}+\beta_{1}D,\quad E\left(\beta\mid D,X\right)=E\left(\beta\mid X\right),\quad\beta\equiv(\beta_{0},\beta_{1})^{\textrm{T}}.
\]
The identification condition is that the conditional second moment
matrix of $(1,D)^{\textrm{T}}$ given $X$ is nonsingular with probability
one, which is the same as
\begin{equation}
\text{var}(D\mid X)=P(X)\{1-P(X)\}>0,\quad P(X)\equiv\Pr(D=1\mid X),\label{eq:binaryT}
\end{equation}
with probability one, where $P(X)$ is the propensity score. Here
we can see that the identification condition is the same as $0<P(X)<1$
with probability one, which is the standard identification condition.

Because $p(D)$ includes an intercept, the
identification condition is the same as nonsingularity of the variance
matrix $\text{var}(D\mid X)$ with probability one. This result generalizes 
(\ref{eq:binaryT}). 

\begin{thm}
$E\{p(D)p(D)^{\textrm{T}}\mid X\}$
is nonsingular with probability one if, and only if, the variance
matrix $\text{var}(D\mid X)$ is nonsingular with probability one.\label{thm:Theorem2} 
\end{thm}

\begin{sloppy}Considerable simplification occurs with mutually exclusive
treatment regimes, which
allows for the formulation of 
an equivalent
condition for nonsingularity of $E\{p(D)p(D)^{\textrm{T}}\mid X\}$ solely in terms of the generalized propensity
scores (\citealp{Imbens:2000}). This result generalizes 
the standard identification condition for binary $D$. 
\par\end{sloppy}

\begin{thm}
With mutually exclusive treatment regimes, $E\{p(D)p(D)^{\textrm{T}}\mid X\}$
is nonsingular with probability one if, and only if, $\Pr\{D(t)=1\mid X\}>0$
for each $t\in\mathcal{T}$ and
\[
\Sigma_{s=1}^{T}\Pr\{D(s)=1\mid X\}<1,
\]
with probability one.\label{thm:Theorem3} 
\end{thm}

For mutually exclusive treatment regimes, the two standard
assumptions for identification 
are common support, i.e., $\Pr(D=0_{T}\mid X)>0$ and $\Pr\{D(t)=1\mid X\}>0$
with probability one for each $t\in\mathcal{T}$, and conditional
independence, i.e.,
\begin{equation}
Y(t)\perp D \mid X, \quad (t = 0,1,\ldots,T);\label{eq:CIA_Potentialoutcomes}
\end{equation}
cf., for instance,  \citet[pp. 190\textendash 192]{Frolich:2004} for a review.
By conditional probabilities adding up to unity, 
Theorem \ref{thm:Theorem3} shows that common support is equivalent to conditional
nonsingularity, and hence is necessary as well as sufficient for identification under Assumption \ref{ass:Assumption1}. It follows that, provided common support holds, identification only requires conditional mean independence, and assumption (\ref{eq:CIA_Potentialoutcomes}) is not necessary. 

For non-mutually exclusive treatment regimes, conditional independence assumption (\ref{eq:CIA_Potentialoutcomes}) 
and common support are not jointly necessary either. In that case, the marginal support $\mathcal{D}$ of $D$ has cardinality $\widetilde{T}>T$. Suppose there exist both a subset $\widetilde{\mathcal{D}}\subset\mathcal{D}$
of cardinality $T$ such that $E\{\mathbb{1}(D\in\widetilde{\mathcal{D}})p(D)p(D)^{\textrm{T}}\mid X\}$
is nonsingular with probability one, and a value $\overline{d}\in\mathcal{D}\backslash\widetilde{\mathcal{D}}$
such that $\Pr(D=\overline{d}\mid X)=0$ with positive probability. Then
common support does not hold but $E\{p(D)p(D)^{\textrm{T}}\mid X\}$ is nonsingular
with probability one, and hence $\mu(D)$ is identified under Assumption \ref{ass:Assumption1}. 
Therefore, 
Theorems  \ref{thm:Theorem1} and  \ref{thm:Theorem3} together 
establish that conditional independence assumption (\ref{eq:CIA_Potentialoutcomes}) and common support are not jointly necessary  for identification, for either type of treatment regime.

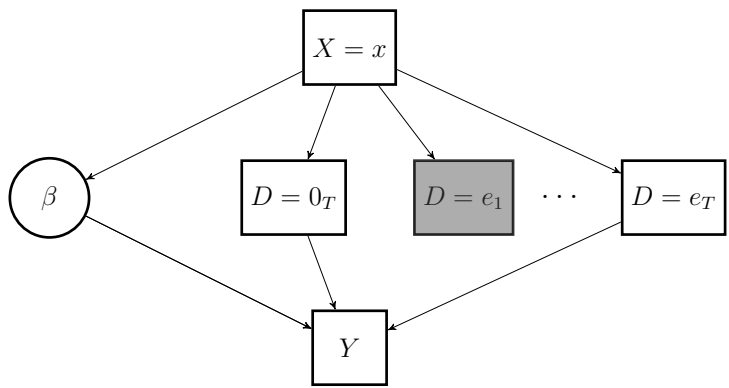
\begin{figure}[t]
\vspace*{0pt}

\begin{centering}
\scalebox{.85}{ 	
\begin{tikzpicture}[->,>=stealth']
	
\node[sq] (X) {$X=x$};
\node[est] (B) [below left = 1.3cm and 3.5cm of X ] {$\beta$};
\node[sq] (D0) [below right = 1.15cm and -2.45cm of X ] {$D=0_{T}$};
\node[sqf] (D1) [below right = 1.15cm and .25cm of X ] {$D=e_{1}$}; 	
\node[myempty] (D3) [below right = 1.4cm and 2.2cm of X ] {\large $\ldots$}; 
\node[sq] (DT) [below right = 1.15cm and 3.5 cm of X ] {$D=e_{T}$}; 	
\node[sq] (Y) [below =  3.5cm of X] {$Y$};
	
\path   
(X) edge node {} (B) 	
(B) edge node {} (Y) 	

(X) edge node {} (D0)
(X) edge node {} (D1)
(X) edge node {} (DT) 	
(B) edge node {} (Y) 			
(D0) edge node {} (Y) 			
(DT) edge node {} (Y);
\end{tikzpicture}} 	
\par\end{centering}

\caption{A directed acyclic graph representation of lack of identification caused by conditional singularity: treatments are mutually exclusive and treatment $1$ occurs with zero probability given $X=x$, and hence the corresponding treatment effect is not identified.} 	
\label{fig:dag2} 
\end{figure}

Theorem \ref{thm:Theorem3} also clarifies the role played by conditional nonsingularity in identification. For mutually exclusive treatment regimes 
$\Pr\{D(t)=1\mid X\}= \Pr(D=e_{t}\mid X)$, $t\in\mathcal{T}$, and hence conditional singularity on a set with positive probability means that at least one of the events $\{D=0_{T}\}$,$\{D=e_{1}\}$,$\ldots$,$\{D=e_{T}\}$, has probability zero conditional on $X$ on that set. Therefore, in this instance, failure of identification occurs because common support 
does not hold. 
Using a directed acyclic graph (\citealp{Pearl2009}), Figure \ref{fig:dag2} illustrates this failure with the event $\{D=e_{1}\}$ having probability zero given $X=x$, for almost every $x$ in a set with positive probability. 
The absence of an arrow between this event 
and $Y$ reflects failure of identification.

\subsection{Extensions}

Our identification results are useful for the analysis of
other interesting objects. When Assumption \ref{ass:Assumption1}
is strengthened to conditional independence
\begin{equation}
\beta\,\bot\,D\mid X,\label{eq:CIA}
\end{equation}
conditional nonsingularity is also sufficient for identification of
distributional and quantile treatment effects. Define the distribution
structural function $G(y,d)$ and, when $Y$ is continuous, the quantile
structural function $Q(\tau,d)$ by 
\[
G(y,d)\equiv\text{Pr}\{p(d)^{\textrm{T}}\beta\leq y\},\quad Q(\tau,d)\equiv\tau^{\text{th}}\text{ quantile of }p(d)^{\textrm{T}}\beta,
\]
where $d$ is fixed in these expressions. 
Distributional and quantile
treatment effects are formed as $G(y,e_{t})-G(y,0_{T})$ and $Q(\tau,e_{t})-Q(\tau,0_{T})$,
respectively, for each $t\in\mathcal{T}$, and pairwise distributional
and quantile treatment comparisons as $G(y,e_{t})-G(y,e_{s})$ and
$Q(\tau,e_{t})-Q(\tau,e_{s})$, respectively, for any $s,t\in\mathcal{T}$,
$s\neq t$.

With mutually exclusive treatment regimes, by Theorem \ref{thm:Theorem3}
the conditional support of $D$ given $X$ coincides with the marginal
support of $D$, and hence the conditional support of $X$ given $D$
coincides with the marginal support of $X$, with probability one. Therefore, by \citet[p. 1489]{ImbensNewey 2009}
conditional nonsingularity and conditional independence property (\ref{eq:CIA}) together imply identification
of $G(Y,D)$ and, when $Y$ is continuous, also of $Q(\tau,D)$, from
\[
G(Y,D)=\int F_{Y\mid DX}(Y\mid D,X=x)F_{X}(dx),\quad Q(\tau,D)=G^{-1}(\tau,D),
\]
where $F_{Y\mid DX}(Y\mid D,X)$ and $F_{X}(X)$ are the cumulative distribution functions of $Y$ given $(D,X)$ and of $X$, respectively, and $\tau\mapsto G^{-1}(\tau,D)$ denotes the inverse function of
$y\mapsto G(y,D)$.

With non-mutually exclusive treatment  regimes,
conditional nonsingularity need not coincide with common support. Identification without common support can nonetheless be achieved under
additional restrictions imposed on model (\ref{eq:g(d,b)}). 
When $Y$ is continuous and letting $Q_{\beta_{t}\mid DX}(u\mid D,X)$ denote the conditional quantile function of $\beta_{t}$ given $(D,X)$, $u\in(0,1)$, an example of sufficient model restrictions 
is that unobserved heterogeneity components $\beta_{t}$ 
satisfy conditional independence property (\ref{eq:CIA}) as well as the additional 
scalar heterogeneity restriction
\begin{equation}
\beta_{t}=Q_{\beta_{t}\mid DX}(U\mid D,X),\quad U\mid D,X\sim \textrm{Un}(0,1), \quad (t = 0,1,\ldots,T),\label{eq:scalar_heterogeneity}
\end{equation}
where the unobservable $U$ is the same for each $\beta_{t}$. 
The control quantile regression function of $Y$ given $(D,X)$ then takes the linear form
\begin{eqnarray*}
Q_{Y\mid DX}(U\mid D,X)&=&p(D)^{\textrm{T}}q_{U}(X),\\ q_{U}(X) &\equiv& \{Q_{\beta_{0}\mid X}(U\mid X),Q_{\beta_{1}\mid X}(U\mid X),\dots,Q_{\beta_{T}\mid X}(U\mid X)\}^{\textrm{T}},
\end{eqnarray*}
\begin{sloppy}by strict monotonicity of $u\mapsto Q_{\beta_{0}\mid DX}(u\mid D,X)+\sum_{t=1}^{T}D(t)Q_{\beta_{t}\mid DX}(u\mid D,X)$.
Conditional nonsingularity implies identification of $q_{U}(X)$, 
and hence of $Q_{Y\mid DX}(U\mid D,X)$. Since the structural functions $G(Y,D)$ and $Q(\tau,D)$
are known functionals of $F_{Y\mid DX}(Y\mid D,X)$, the relation 
\[
F_{Y\mid DX}(Y\mid D,X)=\int_{0}^{1}\mathbb{1}\{Q_{Y\mid DX}(u\mid D,X)\leq Y\}du
\]
implies identification of distributional and quantile treatment effects
(\citealp{NeweyStouli:2021}).

Theorem \ref{thm:Theorem4} summarizes the above discussion
of the role of conditional nonsingularity in identification of distributional
and quantile treatment effects.
\begin{thm}
Suppose that conditional independence property (\ref{eq:CIA})
holds and $E\{p(D)p(D)^{\textrm{T}}\mid X\}$ is nonsingular with probability one. The
following hold: (i) with mutually exclusive treatment regimes, $G(Y,D)$
and, when $Y$ is continuous, also $Q(\tau,D)$, $\tau\in(0,1)$, are identified;
(ii) with non-mutually exclusive treatment  regimes and continuous
outcome $Y$, if the scalar heterogeneity restriction (\ref{eq:scalar_heterogeneity})
holds and $\sup_{u\in(0,1)} E\{|| q_{u}(X) ||^{2}\}<\infty$, then  $G(Y,D)$ and $Q(\tau,D)$, $\tau\in(0,1)$, are identified.\label{thm:Theorem4}
\end{thm}
\end{sloppy}
Other objects of interest include treatment effects on the
treated. For some specified treatment $s\in\mathcal{T}$, average effects are formed using the average structural function for
the treated, $\mu(D,e_{s})\equiv p(D)^{\textrm{T}}E(\beta\mid D=e_{s})$. 
Distributional 
and, when $Y$ is continuous, quantile treatment effects are formed using the distribution and quantile structural functions for the treated,
\begin{eqnarray*}
G(y,d,e_{s})&\equiv&\text{Pr}\{p(d)^{\textrm{T}}\beta\leq y \mid D=e_{s}\},\\Q(\tau,d,e_{s})&\equiv&\tau^{\text{th}}\text{ quantile of }p(d)^{\textrm{T}}\beta\textrm{ given }D=e_{s},
\end{eqnarray*}
respectively, 
where $d$ is fixed in these expressions. These  structural objects are useful for decomposition
and counterfactual analysis (e.g., \citealp{Ao et al:2021}). The
average effect of treatment $t$ on units treated with treatment $s$ is $\mu(e_{t},e_{s})-\mu(0_{T},e_{s})$, and distributional 
and quantile effects of treatment $t$ on units treated with
treatment $s$ are $G(y,e_{t},e_{s})-G(y,0_{T},e_{s})$
and $Q(\tau,e_{t},e_{s})-Q(\tau,0_{T},e_{s})$, respectively. 

Let $\mathcal{X}(s)$ denote the conditional support of $X$ given $D=e_{s}$. The average structural function for the
treated can be expressed as a linear combination of $E\{q_{0}(X)\mid D=e_{s}\}$.
By conditional mean independence and iterated expectations,
\[
p(D)^{\textrm{T}}E\{q_{0}(X)\mid D=e_{s}\}=p(D)^{\textrm{T}}E\{E(\beta\mid X,D=e_{s})\mid D=e_{s}\}=\mu(D,e_{s}),
\]
and hence $\mu(D,e_{s})$ is identified if $q_{0}(X)$ is identified on $\mathcal{X}(s)$.
Thus, for average treatment effects on the treated, the identification
condition becomes nonsingularity of $E\{p(D)p(D)^{\textrm{T}}\mid X=x\}$ for almost every $x$ in the set $\mathcal{X}(s)$. 
This conditional nonsingularity condition is
also necessary for identification of $\mu(D,e_{s})$ under the additional condition that $E\{p(D)p(D)^{\textrm{T}}\}$
is nonsingular.

\begin{thm}
Suppose that Assumption \ref{ass:Assumption1} holds, 
$E\{p(D)p(D)^{\textrm{T}}\}$ is nonsingular, and $\sup_{x\in \mathcal{X}(s)} E(\left\Vert \beta\right\Vert ^{2}\mid X=x)<\infty$ for some specified $s\in\mathcal{T}$ such that $\Pr\{\mathcal{X}(s)\}>0$.
Then: $E\{p(D)p(D)^{\textrm{T}}\mid X=x\}$ is nonsingular for almost every $x\in\mathcal{X}(s)$ 
if, and only if, $\mu(D,e_{s})$ is identified. Furthermore, with mutually exclusive treatment regimes and for almost every $x\in\mathcal{X}(s)$, $E\{p(D)p(D)^{\textrm{T}}\mid X=x\}$
is nonsingular 
if, and only if, $\Pr\{D(t)=1\mid X=x\}>0$
for each $t\in\mathcal{T}$ and $\Sigma_{s=1}^{T}\Pr\{D(s)=1\mid X=x\}<1$. \label{thm:Theorem5}
\end{thm}

If conditional independence property (\ref{eq:CIA}) holds, 
the distribution and, when $Y$ is continuous, quantile structural functions for the
treated also are identified, from
\[
G(Y,D,e_{s})=\int F_{Y| DX}(Y| D,X=x)F_{X| D}(dx| D=e_{s}),\,Q(\tau,D,e_{s})=G^{-1}(\tau,D,e_{s}),
\]
respectively, where $\tau\mapsto G^{-1}(\tau,D,e_{s})$ denotes the inverse function of
$y\mapsto G(y,D,e_{s})$. Here identification only requires the support of $X$ conditional
on $D$ to contain $\mathcal{X}(s)$ with probability one, and hence that the support of $D$ conditional on $X=x$ be the same as the marginal support of $D$ for almost every $x\in\mathcal{X}(s)$. With mutually exclusive treatment regimes, this support condition is equivalent to nonsingularity of $E\{p(D)p(D)^{\textrm{T}}\mid X=x\}$ for almost every $x\in\mathcal{X}(s)$, by Theorem \ref{thm:Theorem5}. Therefore, this conditional nonsingularity condition is sufficient for identification. With non-mutually exclusive treatment regimes, this 
condition is also sufficient for identification of $q_{U}(X)$ on $\mathcal{X}(s)$, and hence of $Q_{Y\mid DX}(U\mid D,X)$ and $F_{Y\mid DX}(Y\mid D,X)$ on $\mathcal{D}\times \mathcal{X}(s)$, when the outcome $Y$ is continuous and the scalar heterogeneity restriction (\ref{eq:scalar_heterogeneity})
holds. 
Thus, results 
analogous to Theorem \ref{thm:Theorem4} hold 
for distribution and quantile treatment effects on the treated. 

\begin{sloppy}
\begin{thm}
Suppose that conditional independence property (\ref{eq:CIA}) holds 
and $E\{p(D)p(D)^{\textrm{T}}| X=x\}$ is nonsingular for almost every $x\in\mathcal{X}(s)$, for some specified $s\in\mathcal{T}$ such that $\Pr\{\mathcal{X}(s)\}>0$. The
following hold: (i) with mutually exclusive treatment regimes, $G(Y,D,e_{s})$
and, when $Y$ is continuous, also $Q(\tau,D,e_{s})$, $\tau\in(0,1)$, are identified;
(ii) with non-mutually exclusive treatment  regimes and continuous
outcome $Y$, if the scalar heterogeneity restriction (\ref{eq:scalar_heterogeneity})
holds and $\sup_{(u,x)\in(0,1)\times \mathcal{X}(s)} E\{|| q_{u}(X) ||^{2}\mid X=x\}<\infty$, then  $G(Y,D,e_{s})$ and $Q(\tau,D,e_{s})$, $\tau\in(0,1)$, are identified.\label{thm:Theorem6}
\end{thm}
\par\end{sloppy}


\section{Discussion}

The heterogeneous coefficients formulation we propose for multiple
treatment effects reveals the central role of the conditional nonsingularity
condition for identification. Because this condition is in principle
testable, establishing that it is also necessary 
demonstrates testability
of identification (e.g., \citealp{Breusch:1986}). With mutually exclusive
treatments, the formulation of the equivalent common support condition in Theorem \ref{thm:Theorem3} 
thus relates testability of identification to the generalized propensity
scores. This is a generalization of the relationship between testability
of identification and the propensity score in the binary treatment
case.

Conditions that are both necessary and sufficient are also important
for the determination of minimal conditions for identification. In an 
unpublished 2004 working paper (cemmap CWP03/04), Wooldridge 
considers a restricted version
of our model 
with $E(D\mid \beta,X)=E(D\mid X)$
and $E\{p(D)p(D)^{\textrm{T}}\mid \beta,X\}=E\{p(D)p(D)^{\textrm{T}}\mid X\}$, and shows that $q_{0}(X)$
is identified if $E\{p(D)p(D)^{\textrm{T}}\mid X\}$ is invertible. The additional conditional
second moments assumption implies that his identification condition
differs from ours. Thus his result and proof do not apply in our setting
which only assumes conditional mean independence $E(\beta\mid D,X)=E(\beta\mid X)$,
and our results show that conditional second moments independence
is not necessary for identification in multiple treatment effect models.
\citet{Graham Pinto 2018} consider a related approach in work independent 
of the 
first version of this paper \citep{NeweyStouli:2018} 
where we derived our identification result (Lemma \ref{lem:Lemma1} in the Appendix). 
The conditional nonsingularity condition we propose is weaker than their identification condition, and we study necessity as well as sufficiency for identification of average treatment effects.

We analyze the role of conditional nonsingularity for identification of multiple treatment effects under the maintained conditional
mean independence Assumption \ref{ass:Assumption1}. Although itself
not testable in general, this assumption is substantially weaker than
the standard conditional independence property (\ref{eq:CIA_Potentialoutcomes}). 
In the general case
of mutually exclusive treatment regimes, the relaxation of conditional
independence afforded by our heterogeneous coefficients approach is the same as the conditional mean independence condition
\begin{equation}
E\{Y(t)\mid D,X\}=E\{Y(t)\mid X\} \quad (t = 0,1,\ldots,T),\label{eq:CMIA}
\end{equation}
because the formulation of Assumption \ref{ass:Assumption1} in terms
of potential outcomes,
\[
E\{Y(0)| D,X\}=E\{Y(0)| X\},\;\; E\{Y(t)-Y(0)| D,X\}=E\{Y(t)-Y(0)| X\},\;\; t\in\mathcal{T},
\]
reveals that Assumption \ref{ass:Assumption1} is equivalent to (\ref{eq:CMIA}). Therefore, our results allow applied researchers to replace unconfoundedness requirement (\ref{eq:CIA_Potentialoutcomes}) for
identification by the weaker condition (\ref{eq:CMIA}) under which conditional nonsingularity
is both necessary and sufficient for identification, thereby improving robustness of empirical studies in nonexperimental settings. 
In particular, conditional mean independence property (\ref{eq:CMIA}) allows for any higher conditional moment of $Y(t)$
to depend on both $D$ and $X$. 
Our identification results 
are thus of general interest
for the vast treatment effects literature (e.g., 
\citealp{Ath Imbens:2017} for a recent literature review) 
and 
complement 
existing results 
on identification of treatment effects.

\newpage

\appendix

\section{Treatment effects modeling with non-mutually exclusive treatment regimes \label{sec:Model}}

For non-mutually exclusive treatment regimes, there are $\widetilde{T}-T \geq 1$ 
combinations of 
treatments in $\mathcal{T}\equiv\{1,\dots,T\}$, 
denoted $\mathcal{C}(s)$ with 
$s\in\{T+1,\ldots,\widetilde{T}\}\equiv\widetilde{\mathcal{T}}$.
For each $s\in\widetilde{\mathcal{T}}$, multiple components of $D$ 
take value one jointly if treatment combination
$\mathcal{C}(s)$ occurs. Define $\widetilde{D}$
a vector of dummy variables $\widetilde{D}(s)$ taking value one for $s\in\mathcal{T}$ if only treatment $s\in\mathcal{T}$ occurs, 
and for $s\in\widetilde{\mathcal{T}}$ if treatment combination $\mathcal{C}(s)$
occurs. A general model that gives
rise to an average structural function of the form (\ref{eq:mu(D)}) 
is
\begin{equation}
Y=p(\widetilde{D})^{\textrm{T}}\widetilde{\beta}=\widetilde{\beta}_{0}+
\sum_{s=1}^{\widetilde{T}}\widetilde{D}(s)\widetilde{\beta}_{s},\quad E(\widetilde{\beta}\mid X)=E(\widetilde{\beta}\mid\widetilde{D},X),\quad \widetilde{\beta}=(\widetilde{\beta}_{0},\widetilde{\beta}_{1},\ldots,\widetilde{\beta}_{\widetilde{T}})^{\textrm{T}},\label{eq:Saturated_model}
\end{equation}
restricted so that heterogeneity satisfies the conditional average
additivity property
\begin{equation}
E(\widetilde{\beta}_{s}\mid X)=\sum_{t=1}^{T}\mathbb{1}\{t\in\mathcal{C}(s)\} E(\widetilde{\beta}_{t}\mid X),\quad s\in\widetilde{\mathcal{T}}.\label{eq:additive effects}
\end{equation}
For $s\in\mathcal{T}$, extending the definition of $\mathcal{C}(s)$ by setting $\mathcal{C}(s)=\{s\}$ 
and also writing $E(\widetilde{\beta}_{s}\mid X)= \sum_{t=1}^{T}\mathbb{1}\{t\in\mathcal{C}(s)\}E(\widetilde{\beta}_{t}\mid X)$, 
the implied control regression function $E(Y\mid \widetilde{D},X)$
for model (\ref{eq:Saturated_model})-(\ref{eq:additive effects}) takes the form
\begin{align*}
p(\widetilde{D})^{\textrm{T}}E(\widetilde{\beta}| X) & =E(\widetilde{\beta}_{0}\mid X)+
\sum_{s=1}^{\widetilde{T}}\widetilde{D}(s)\left[\sum_{t=1}^{T}\mathbb{1}\{t\in\mathcal{C}(s)\} E(\widetilde{\beta}_{t}\mid X)\right] \\
 & =E(\widetilde{\beta}_{0}\mid X)+\sum_{t=1}^{T}\left[ \sum_{s=1}^{\widetilde{T}}\mathbb{1}\{t\in\mathcal{C}(s)\}\widetilde{D}(s)\right] E(\widetilde{\beta}_{t}\mid X)=p(D)^{\textrm{T}}E(\beta| X),
\end{align*}
the control regression function for model (\ref{eq:g(d,b)}) with
$\beta = (\widetilde{\beta}_{0},\widetilde{\beta}_{1},\ldots,\widetilde{\beta}_{T})^{\textrm{T}}$
and $D$ such that $D(t) = \sum_{s=1}^{\widetilde{T}}\mathbb{1}\{t\in\mathcal{C}(s)\}\widetilde{D}(s)$,
$t\in\mathcal{T}$. Therefore, the control regression functions, and hence also the average structural functions, for
model (\ref{eq:g(d,b)}) with non-mutually exclusive treatment regimes and for model (\ref{eq:Saturated_model})-(\ref{eq:additive effects}) coincide.

\section{Proofs\label{sec:Proofs}}

\subsection*{Preliminary result}
\begin{lem}
Suppose that $E(\left\Vert \beta\right\Vert ^{2})<\infty$ and
Assumption \ref{ass:Assumption1} holds. If $E\{p(D)p(D)^{\textrm{T}}\mid X\}$ is nonsingular
with probability one then $q_{0}\left(X\right)$ is identified.\label{lem:Lemma1}
\end{lem}
\begin{proof}
\begin{sloppy}Let $\lambda_{\min}(X)$ denote the smallest eigenvalue of $E\{p(D)p(D)^{\textrm{T}}\mid X\}$.
Suppose that $\overline{q}\left(X\right)\neq q_{0}\left(X\right)$ with
positive probability on a set $\widetilde{\mathcal{X}}$, and note
that $\lambda_{\min}(X)>0$ on $\mathcal{X}$ by assumption. Then
\begin{eqnarray*}
E\left( [ p(D)^{\textrm{T}} \{ \overline{q}(X)-q_{0}(X)\} ]^{2} \right) 
& = & E\left[
\{ \overline{q}(X)-q_{0}(X)\} ^{\textrm{T}}
E\{p(D)p(D)^{\textrm{T}}| X\}\{ \overline{q}(X)-q_{0}(X)\} 
\right]\\
 & \geq & E\left\{
 \left\Vert \overline{q}\left(X\right)-q_{0}\left(X\right)\right\Vert ^{2}\lambda_{\min}\left(X\right)\right\}\\
 & \geq & E\left\{\mathbb{1}(X\in\mathcal{X}\cap\widetilde{\mathcal{X}})\left\Vert \overline{q}\left(X\right)-q_{0}\left(X\right)\right\Vert ^{2}\lambda_{\min}\left(X\right)\right\}.
\end{eqnarray*}
By definition $\Pr(\widetilde{\mathcal{X}})>0$ and $\widetilde{\mathcal{X}}\subseteq\mathcal{X}$
so that $\widetilde{\mathcal{X}}\cap\mathcal{X}=\widetilde{\mathcal{X}}$.
Thus the fact that $\left\Vert \overline{q}\left(X\right)-q_{0}\left(X\right)\right\Vert ^{2}\lambda_{\min}\left(X\right)$
is positive on $\widetilde{\mathcal{X}}\cap\mathcal{X}$ implies 
\[
E\left\{\mathbb{1}(X\in\mathcal{X}\cap\widetilde{\mathcal{X}})\left\Vert \overline{q}\left(X\right)-q_{0}\left(X\right)\right\Vert ^{2}\lambda_{\min}\left(X\right)\right\}>0.
\]
\end{sloppy} We have shown that, for $\overline{q}\left(X\right)\neq q_{0}\left(X\right)$
with positive probability on a set $\widetilde{\mathcal{X}}$, 
\begin{equation*}
E\left( [ p(D)^{\textrm{T}} \{ \overline{q}(X)-q_{0}(X)\} ]^{2} \right) 
>0,
\end{equation*}
which implies $p\left(D\right)^{\textrm{T}}\overline{q}\left(X\right)\neq p\left(D\right)^{\textrm{T}}q_{0}\left(X\right)$.
Therefore, $q_{0}\left(X\right)$ is identified from $E(Y\mid D,X)$.
\end{proof}
%

\subsection*{Proof of Theorem \protect\ref{thm:Theorem1}}

We first show that nonsingularity of $E\{p(D)p(D)^{\textrm{T}}\mid X\}$ with probability
one implies identification of $\mu(D)$. By Lemma \ref{lem:Lemma1},
if $E\{p(D)p(D)^{\textrm{T}}\mid X\}$ is nonsingular with probability one then $q_{0}(X)$
is identified, and hence $E\{q_{0}(X)\}$ also is. By $p(D)$ being
a known function, $p(D)^{\textrm{T}}E\{q_{0}(X)\}=\mu(D)$ is identified.

\begin{sloppy}We now establish that nonsingularity of $E\{p(D)p(D)^{\textrm{T}}\mid X\}$ with probability
one is necessary for identification of $\mu(D)$. It suffices to show 
that singularity of $E\{p(D)p(D)^{\textrm{T}}\mid X\}$ with positive probability implies
that $\mu(D)$ is not identified, i.e., there exists an observationally equivalent $\overline{q}(X)\neq q_{0}(X)$
with positive probability such that $p(D)^{\textrm{T}}E\{\overline{q}(X)\}\neq p(D)^{\textrm{T}}E\{q_{0}(X)\}$
with positive probability. 
By nonsingularity of $E\{p(D)p(D)^{\textrm{T}}\}$ and
linearity of $\mu(D)$, the conclusion holds if, and only if, there exists an observationally equivalent $\overline{q}(X)\neq q_{0}(X)$
with positive probability such that $E\{\overline{q}(X)\}\neq E\{q_{0}(X)\}$.\par\end{sloppy}

Suppose that  $E\{p(D)p(D)^{\textrm{T}}\mid X\}$ is singular with positive probability and let $\Delta(X)$ be such that $E\{p(D)p(D)^{\textrm{T}}\mid X\}\Delta(X)=0$. We have that 
$\Delta(X)\neq 0$ on a set
$\widetilde{\mathcal{X}}$ with $\Pr(\widetilde{\mathcal{X}})>0$. For $J=T+1$, define $\widetilde{\mathcal{X}}_{j}=\{x\in\widetilde{\mathcal{X}}:\Delta_{j}(x)\neq0\}$,
$j\in\{1,\dots,J\}$. Then $\cup_{j=1}^{J}\widetilde{\mathcal{X}}_{j}=\{x\in\widetilde{\mathcal{X}}:\Delta(x)\neq0\}=\widetilde{\mathcal{X}}$.
Hence
\[
0<\Pr(\widetilde{\mathcal{X}})=\Pr(\cup_{j=1}^{J}\widetilde{\mathcal{X}}_{j})\leq\sum_{j=1}^{J}\Pr(\widetilde{\mathcal{X}}_{j}),
\]
which implies that $\Pr(\widetilde{\mathcal{X}}_{j^{*}})>0$ for some
$j^{*}\in\{1,\dots,J\}$. 

Set $\widetilde{\Delta}(x)=\Delta(x)$ for
$x\in\widetilde{\mathcal{X}}_{j^{*}}$, and $\widetilde{\Delta}(x)=0$
otherwise. 
By construction $\widetilde{\Delta}_{j^{*}}(X)\neq0$, and letting
\[
\widetilde{\widetilde{\Delta}}(X)=\textrm{sign}\{\widetilde{\Delta}_{j^{*}}(X)\}\frac{\widetilde{\Delta}(X)}{||\widetilde{\Delta}(X)||},
\]
we have that $\widetilde{\widetilde{\Delta}}_{j^{*}}(X)>0$ on $\widetilde{\mathcal{X}}_{j^{*}}$ and $||\widetilde{\widetilde{\Delta}}(X)||=1$,
and hence $E\{||\widetilde{\widetilde{\Delta}}(X)||\}<\infty$
and $E\{\widetilde{\widetilde{\Delta}}_{j^{*}}(X)\}\neq0$. Therefore
$E\{\widetilde{\widetilde{\Delta}}(X)\}\neq0$, which implies that $E\{q_{0}(X)+\widetilde{\widetilde{\Delta}}(X)\} \neq E\{q_{0}(X)\}$.
The result follows. \qed

\subsection*{Proof of Theorem \protect\ref{thm:Theorem2}}

The matrix $E\{p(D)p(D)^{\textrm{T}}\mid X\}$ is of the form 
\begin{equation}
E\{p(D)p(D)^{\textrm{T}}\mid X\}=\begin{bmatrix}1 & E(D^{\textrm{T}}\mid X)\\
E(D\mid X) & E(DD^{\textrm{T}}\mid X)
\end{bmatrix},\label{eq:principal-2}
\end{equation}
and is positive definite if, and only if, the Schur complement of
$1$ in (\ref{eq:principal-2}) is positive definite (\citealp[Appendix A.5.5.]{Boyd Vandenberghe:2004}),
i.e., if, and only if,
\[
E(DD^{\textrm{T}}\mid X)-E(D\mid X)E(D^{\textrm{T}}\mid X)=\text{var}(D\mid X),
\]
is positive definite with probability one, as claimed.\qed

\subsection*{Proof of Theorem \protect\ref{thm:Theorem3}}

Suppose that the matrix  $E\{p(D)p(D)^{\textrm{T}}\mid X\}$ is nonsingular
with probability one. For mutually exclusive treatment regimes, $D\in\{0_{T},\{e_{t}\}_{t\in\mathcal{T}}\}$
and hence $E\{p(D)p(D)^{\textrm{T}}\mid X\}$ is of the form 
\[
E\{p(D)p(D)^{\textrm{T}}| X\}=\left\{ p(0_{T})p(0_{T})^{\textrm{T}}\right\} \times\Pr(D=0_{T}| X)+\sum_{t=1}^{T}\left\{ p(e_{t})p(e_{t})^{\textrm{T}}\right\} \times\Pr(D=e_{t}| X),
\]
a sum of $T+1$ rank one $(T+1)\times(T+1)$ distinct matrices which
is singular with positive probability if either $\Pr(D=0_{T}\mid X)=0$
or $\Pr(D=e_{t}\mid X)=0$ for some $t\in\mathcal{T}$ with positive probability.
For mutually exclusive treatment regimes
\[
\Pr(D=e_{t}\mid X)=\Pr\{D(t)=1\mid X\},\quad t\in\mathcal{T},
\]
and hence if either $\Pr(D=0_{T}\mid X)=0$ or $\Pr\{D(t)=1\mid X\}=0$ for
some $t\in\mathcal{T}$ with positive probability, then $E\{p(D)p(D)^{\textrm{T}}\mid X\}$
is singular with positive probability. Therefore, nonsingularity of
$E\{p(D)p(D)^{\textrm{T}}\mid X\}$ with probability one implies that $\Pr(D=0_{T}\mid X)>0$
and  $\Pr\{D(t)=1\mid X\}>0$ for each $t\in\mathcal{T}$ with probability
one. Since conditional probabilities add up to unity with probability one, for mutually exclusive treatment regimes
\[
\Pr(D=0_{T}\mid X)+\sum_{t=1}^{T}\Pr\{D(t)=1\mid X\}=1
\]
with probability one, and we
have shown that $\Pr\{D(t)=1\mid X\}>0$ for each $t\in\mathcal{T}$ and
$\Sigma_{s=1}^{T}\Pr\{D(s)=1\mid X\}<1$, with probability one.

We show the converse result. Assume that $\Pr\{D(t)=1\mid X\}>0$
for each $t\in\mathcal{T}$ and $\Sigma_{s=1}^{T}\Pr\{D(s)=1\mid X\}<1$,
with probability one. For a vector $w\in\mathbb{R}^{T}$, let $\text{diag}(w)$
denote the $T\times T$ diagonal matrix with diagonal elements $w_{1},\dots,w_{T}$.
For mutually exclusive treatments, the matrix $E\{p(D)p(D)^{\textrm{T}}\mid X\}$ is
also of the form 
\begin{equation}
E\{p(D)p(D)^{\textrm{T}}\mid X\}=\begin{bmatrix}1 & E(D^{\textrm{T}}\mid X)\\
E(D\mid X) & \text{diag}\{E(D\mid X)\}
\end{bmatrix}.\label{eq:Blockmat}
\end{equation}
The matrix $\text{diag}\{E(D\mid X)\}$ has diagonal elements $E\{D(t)\mid X\}=\Pr\{D(t)=1\mid X\}>0$
for each $t\in\mathcal{T}$, by assumption, and hence is positive
definite and invertible.

By assumption $\Sigma_{s=1}^{T}\Pr\{D(s)=1\mid X\}<1$, and hence
\begin{align*}
0<1-\Sigma_{s=1}^{T}\Pr\{D(s)=1\mid X\} & =1-\Sigma_{s=1}^{T}E\{D(s)\mid X\}\\
 & =1-E(D^{\textrm{T}}\mid X)\text{diag}\{E(D\mid X)\}^{-1}E(D\mid X).
\end{align*}
Thus the Schur complement of $\text{diag}\{E(D\mid X)\}$ in (\ref{eq:Blockmat})
is positive definite, and hence $E\{p(D)p(D)^{\textrm{T}}\mid X\}$ is positive definite
(\citealp[Appendix A.5.5.]{Boyd Vandenberghe:2004}). Therefore, $E\{p(D)p(D)^{\textrm{T}}\mid X\}$
is nonsingular with probability one, as claimed.\qed


\subsection*{Proof of Theorem \protect\ref{thm:Theorem4}}

For mutually exclusive treatments, result (i) follows from equivalence between conditional nonsingularity and common support and the argument in the main text. For non-mutually exclusive treatments, $q_{u}(X)$ is identified for each $u\in (0,1)$ by an argument similar to the proof of Lemma \ref{lem:Lemma1}, upon substituting $q_{u}(X)$ for $q_{0}(X)$. Result (ii) then follows from the argument in the main text.\qed

\subsection*{Proof of Theorem \protect\ref{thm:Theorem5}}

The proof is similar to the proofs of Theorems \ref{thm:Theorem1} and \ref{thm:Theorem3} and hence is omitted.\qed

\subsection*{Proof of Theorem \protect\ref{thm:Theorem6}}

For mutually exclusive treatments, by Theorem \ref{thm:Theorem5} conditional nonsingularity on $\mathcal{X}(s)$ is equivalent to the support of $D$ conditional on $X=x$ being the same as the marginal support of $D$  for almost every $x\in\mathcal{X}(s)$. Hence, the support of $X$ conditional on $D$ contains $\mathcal{X}(s)$ with probability one. Result (i) then follows from the argument in the main text.

For non-mutually exclusive treatments, $q_{u}(X)$ is identified on $\mathcal{X}(s)$ for each $u\in (0,1)$ by an argument similar to the proof of Lemma \ref{lem:Lemma1}, upon substituting $q_{u}(X)$ for $q_{0}(X)$, letting $\overline{q}\left(X\right)\neq q_{u}\left(X\right)$  on a set with
positive probability $\widetilde{\mathcal{X}}\subseteq \mathcal{X}(s)$. 
Result (ii) then follows from the argument in the main text.\qed

\end{document}